\documentclass[preprint,review,3p,12pt]{elsarticle}

\usepackage{amsmath}
\usepackage{graphics}

\journal{Physica A}

\begin{document}

\begin{frontmatter}

\title{Scaling laws in the dynamics of crime growth rate}
\author{Luiz G. A. Alves}
\ead{gustavoandradealves@gmail.com}
\author{Haroldo V. Ribeiro}
\author{Renio S. Mendes}
\address{Departamento de F\'isica and National Institute of Science and Technology for Complex Systems, Universidade Estadual de Maring\'a, Av. Colombo 5790, 87020-900, Maring\'a, PR, Brazil}

\begin{abstract}
The increasing number of crimes in areas with large concentrations of people have made cities one of the main sources of violence. Understanding characteristics of how crime rate expands and its relations with the cities size goes beyond an academic question, being a central issue for contemporary society. Here, we characterize and analyze quantitative aspects of murders in the period from 1980 to 2009 in Brazilian cities. We find that the distribution of the annual, biannual and triannual logarithmic homicide growth rates exhibit the same functional form for distinct scales, that is, a scale invariant behavior. We also identify asymptotic  power-law decay relations between the standard deviations of these three growth rates and the initial size. Further, we discuss similarities with complex organizations.
\end{abstract}

\end{frontmatter}

\section{Introduction}

Methods and techniques inspired by statistical physics have been shown to be useful in the search for hidden features in social systems \cite{castellano,boccara}. Particularly, it has been remarked that social systems can exhibit universal properties similarly to thermodynamical systems at the criticality. For example, scaling laws have been reported  in  scientific research \cite{havemann,picoli}, biological systems \cite{keitt,labra}, economics \cite{stanley,prl,fu} and religious  \cite{ausloos,sergio} activities. In addition, there is some evidence about the relation between urban metrics and the population size, where non-linearities are explicitly manifested by power-laws \cite{pnas,epjb,nature,plos,lievano}. More recently, phase transition was also found in a model for criminality \cite{gordon,iglesias}.

Crime is one of the major concerns of contemporary society and, therefore, there is a great interest in understanding features of its organization and dynamics. We are living in a period when most people live in cities \cite{crane}. The increasing concentration of people in urban areas entails both opportunities and challenges \cite{kates}. In particular, cities have become one of the principal sources of problems such as pollution, spread of disease and crime \cite{pnas,glaeser}. Studies on this last subject involve many areas of knowledge going from  human sciences \cite{glaeser,becker,crane2,glaeser2,ehrlich,levitt,kennedy,messner,blumstein,tedesco,levitt2,keizer,sachsida} to exact sciences \cite{pnas,epjb,nature,plos,lievano,gordon,iglesias,ninob,cook,nuno,short,gordon2,short2,short3}. In particular, the economic relevance of social issues of crime has been discussed \cite{becker}. It has also been pointed out that social interactions may explain the large variance in crime in cities with different concentration of people \cite{glaeser2}. However, despite the broad investigations and the social relevance of criminality in connection with urbanization, our understanding of universality and pervasiveness of patterns in this subject remains limited.  In this work, we apply a statistical approach to investigate patterns and growth characteristics of homicide.

In a quantitative manner, our study provides insights on how the growth of crime and the size of cities are related.
More precisely, we study homicide growth rates in Brazilian cities based on data from 1980 to 2009 (section 2), focusing on scaling laws related to probability distributions and standard deviations. We investigate homicide growth rates of cities aiming to understand the mechanisms that govern criminality (section 3). We show that our results have a similar formal structure to those ones found in complex systems such as scientific research, biological systems, economic and religious activities, fact that put the universality of our findings in evidence (section 4). Motivated by this universality, we indicate a scenario to our results in terms of stochastic models proposed in the context of complex organizations (section 4). Our conclusions are in section 5.

\section{Data Presentation}

For the study of statistical properties of crime, we consider a database containing the annual number of homicides in all Brazilian cities spanning the years from 1980 to 2009, obtained from the database of vital statistics of DATASUS \cite{datasus}. The annual population of cities was obtained from the demographic and socio-economic database of DATASUS \cite{datasus}. {In this last database, the years 1980, 1991, 1996 and 2000 report the population number obtained by the population census conducted by the IBGE \cite{ibge}, while all the other years are actually approximated values of the population number estimated  by IBGE agency. Due to this fact, our analysis will be mainly focused on the homicides database.}

\begin{figure}[!b]
\centering
\includegraphics[scale=0.8]{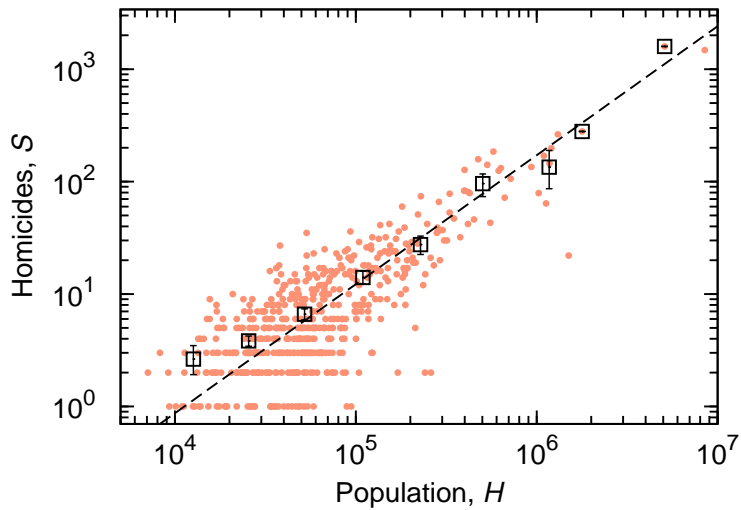}
\caption{The red dots are the data of the number of homicides versus the population for 528 selected cities in 1980 (on a log-log scale). The black squares are the average values of the data binned by window. The black dashed line is a linear fit of the main trend of the data on log-log scale, where we find an exponent close to $1.15$. The error bars are 95\% confidence intervals obtained via bootstrapping \cite{efron}. }
\label{fig:crimepop}
\end{figure}

We selected 528 cities from this set, which present a significant number of homicides (at least one per year) in the period from 1980 to 2009. They are about $10\%$ of Brazilian cities but represent approximately $79\%$ of the total number of homicides in Brazil in the period considered. {Moreover, the average percentage of the population of the country living in these cities during this period is about $58\%$. An illustration of our database is given in Fig. \ref{fig:crimepop}.} {In this figure, a typical scaling law can be observed if we consider only the big cities (population greater than 53.465). We find an exponent very close to those found in other studies on urban metrics and crime \cite{pnas,epjb,nature,plos,lievano}. However, if we take the 528 cities into account the exponent is approximately one. For years subsequent to 1980, the behavior of the exponents is similar.}

\begin{figure}[!b]
\centering
\includegraphics[scale=0.65]{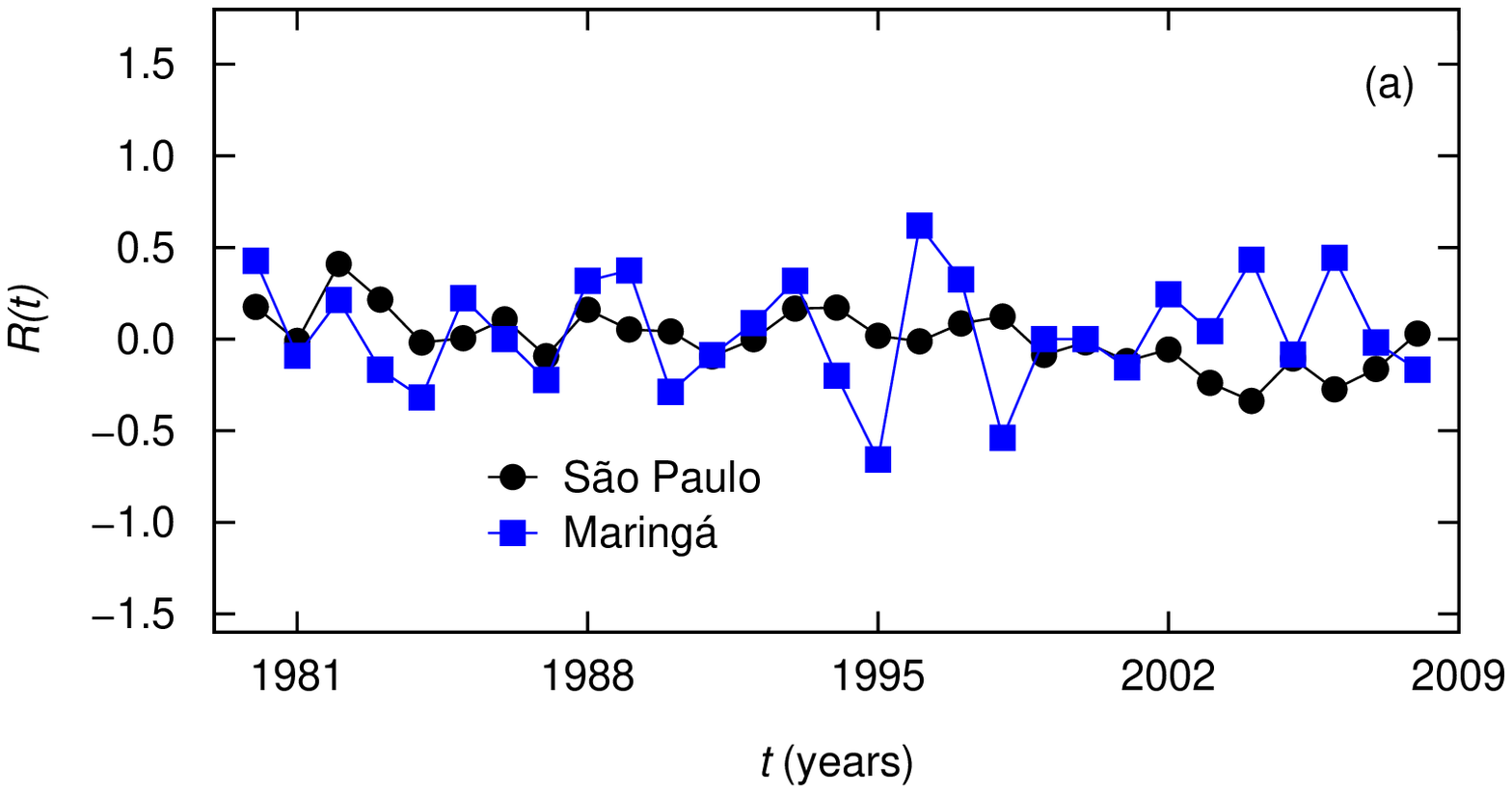}
\includegraphics[scale=0.65]{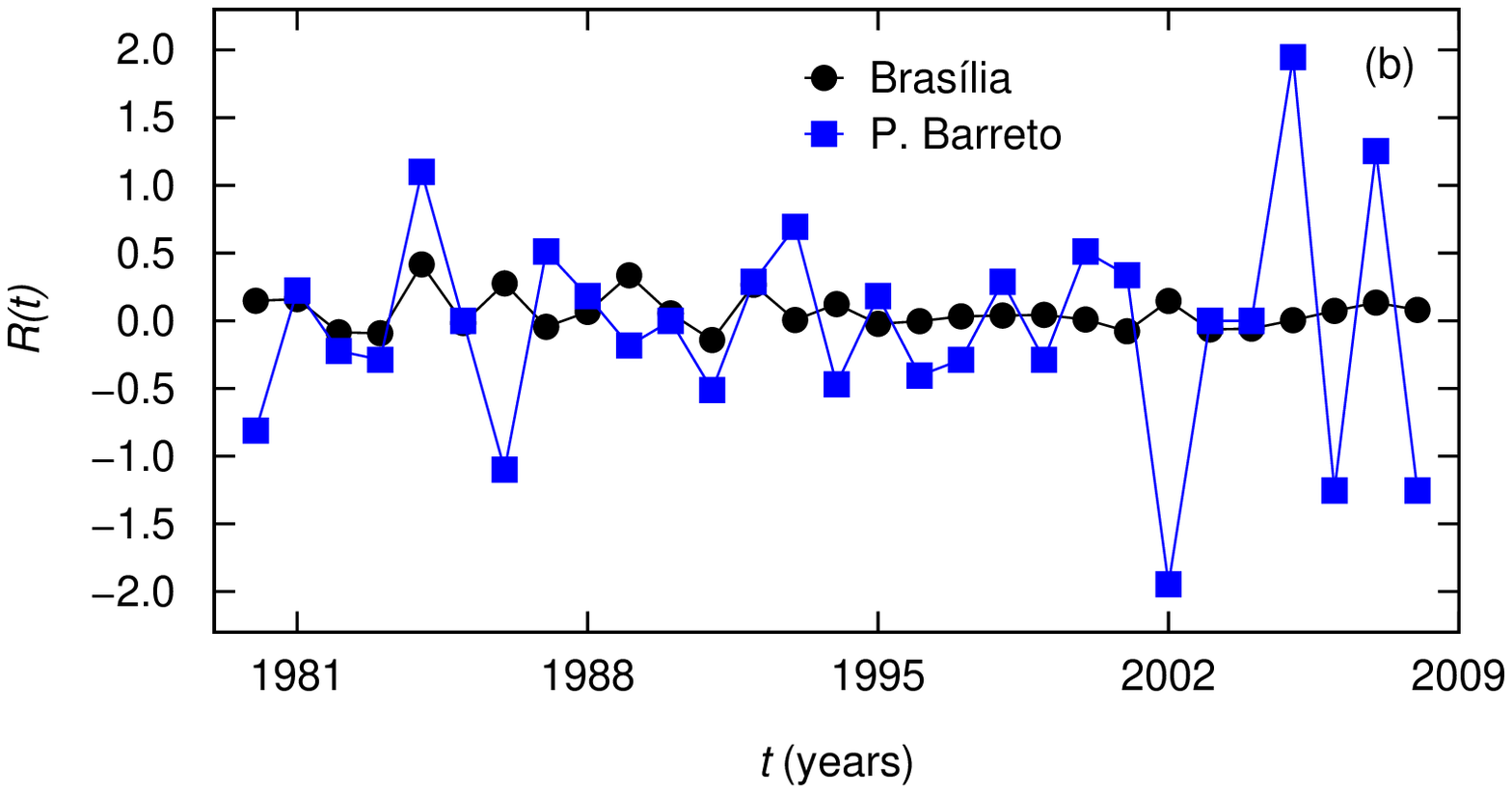}
\caption{Examples of $R(t)$ given by equation (\ref{eq:1}) for some cities in the period from 1980 to 2009. (a) S\~ao Paulo ($S=1480$) and  Maring\'a ($S=15$). (b) Bras\'ilia ($S=145$) and  Pereira Barreto ($S=9$). Observe that the fluctuations in $R(t)$ are generally larger in cities with small number of homicides in 1980, $S$.}
\label{fig:evolution}
\end{figure}

In terms of the total number of homicides $S(t)$ in the year $t$ in a given city, the annual, biannual and triannual (logarithmic) growth rates are defined as
\begin{equation}
R_n(t)=\ln \left[\frac{S(t+n)}{S(t)}\right],
\label{eq:1}
\end{equation}
with $n=1$, $2$ and $3$, respectively. To simplify the notation, we omit the sub-index $n$ when referring to annual growth rates, this is, we employ $R(t)$ to represent $R_1 (t)$. Examples of the temporal evolution of $R(t)$ for some cities are shown in Fig. \ref{fig:evolution}. They illustrate the presence of fluctuations in the homicide growth rate $R(t)$. This figure also exemplifies that the fluctuations of $R(t)$ are generally larger in small towns than in  bigger urban centers. {This is an expected result that motivates our analysis in the standard deviation in function of the city size, since fluctuations are larger in small systems than in bigger systems, as found in different contexts \cite{havemann,picoli,keitt,labra,stanley,prl,fu,sergio}.}

\section{Results}

Even without a detailed description of the system, studies on complex systems have shown that scaling properties of fluctuations in macroscopic quantities can lead to a better understanding of some features of these variables. In  this direction, a variety of complex organizations has been investigated and it has been shown that, despite the differences, they all show non-Gaussian tent-shaped distributions in their annual growth rates with a mean power-law behavior in its standard deviation \cite{havemann,picoli,keitt,labra,stanley,prl,fu,sergio}. These studies have suggested that there are some universal mechanisms, independent of the particular details, governing the fluctuations on growth rates of complex organizations.  These facts motivate us to investigate aspects of the crime growth focusing on these statistical properties.

\begin{figure}[!h]
\centering
\includegraphics[scale=0.64]{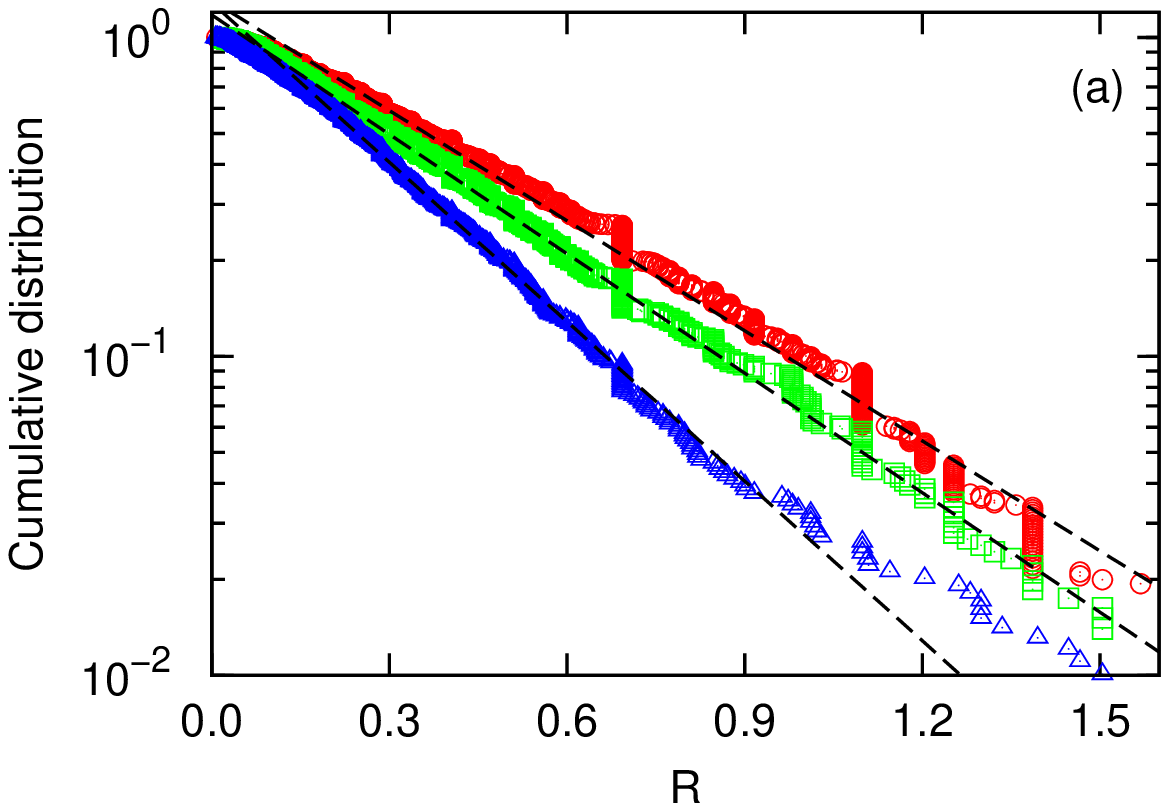}
\includegraphics[scale=0.64]{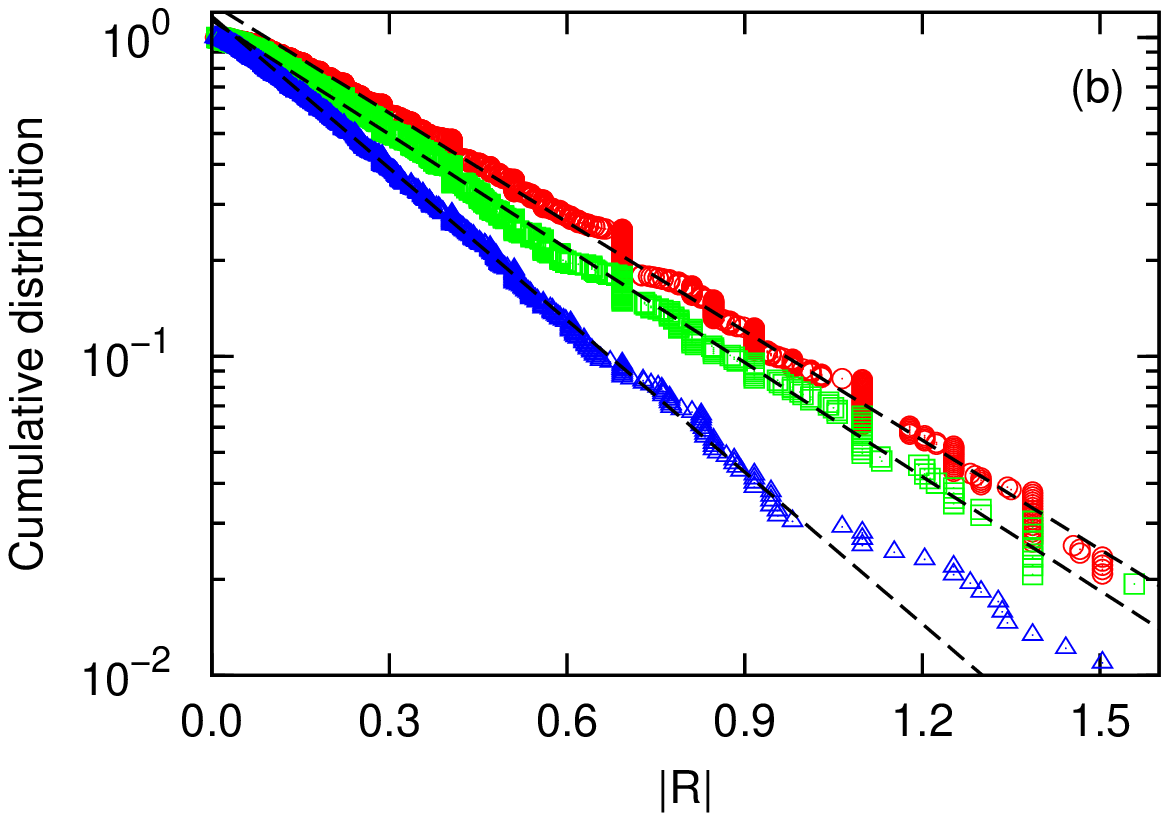}
\caption{Cumulative distributions $p_c(R|S)$ for some intervals of initial number of homicides $S$: red circles ($5 \leq S \leq 8$), green squares ($9 \leq S \leq 13$) and blue triangles ($14 \leq S \leq 23$) {on a mono-log scale}. The dashed lines are Laplace distribution fits. (a) Positive values of $R$ and (b) {module of the negative values of $R$.}}
\label{fig:distribution}
\end{figure}

We first consider the conditional distribution of $R(t)$ by taking some ranges of initial values of $S(t)$ into account. In an attempt to reduce effects of statistical fluctuations, we employ cumulative distribution functions (CDFs). {To use this method, we separate the growth rates $R(t)$ in positive and negative values.} For instance, for positive $R(t)$ we have the CDF
\begin{equation}
p_c(R|S)=\int_R^\infty p(R'|S)\, dR',
\label{eq:2}
\end{equation}
where $p(R'|S)$ is the conditional probability density function. {Figure \ref{fig:distribution} shows examples of empirical CDFs by considering three ranges for the values of $S$ among the set of possible values}, where we find a good agreement between the empirical data and the family of Laplace distributions
\begin{equation}
p(R|S)=\frac{1}{\sqrt{2} \; \sigma}\; \exp \left(-\frac{\sqrt{2}\; |R-\mu |}{\sigma}\right)\,,
\label{eq:3}
\end{equation}
with $\mu$ being the mean value and $\sigma$ being the standard deviation of $R$ {within each selected interval of $S$.}

Scaling behavior can be graphically represented considering an appropriate change in the scale of the independent and dependent variables. If scaling holds, the data for a variety of initial values should collapse upon a single curve. For instance, we note from equation (\ref{eq:3}) that 
\begin{equation}
r(t)=\frac{R(t)-\mu}{\sigma}
\label{eq:4}
\end{equation}
is the scaling variable, where $\mu$ is the mean value and $\sigma$ is the standard deviation of $R$ {within each interval of $S$. This changing of variable leads us to}
\begin{equation}
p(r|S)=\frac{1}{\sqrt{2}}\; \exp \left(-\sqrt{2}\; |r| \right) ,
\label{eq:5}
\end{equation}
where $p(r|S)$ is the probability distribution of the scaled variable $r$, which has no parameters.

\begin{figure}[!h]
\centering
\includegraphics[scale=0.64]{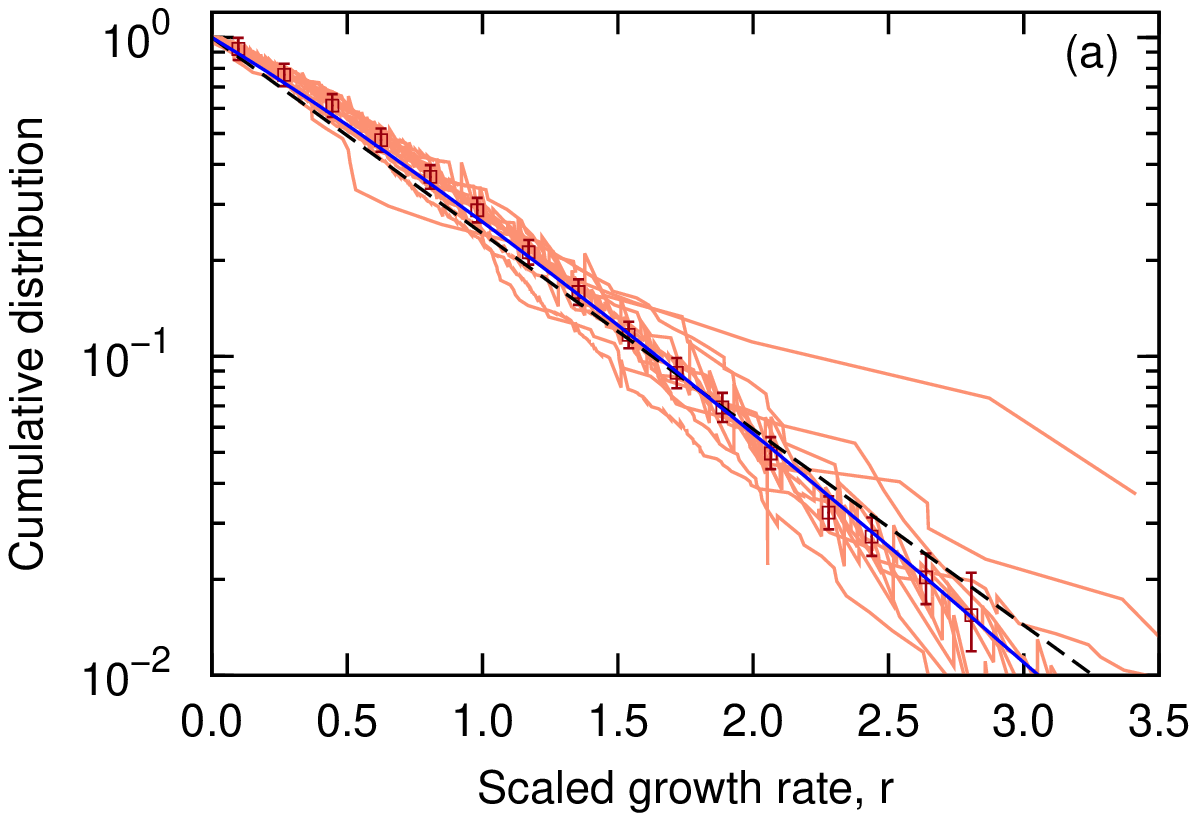}
\includegraphics[scale=0.64]{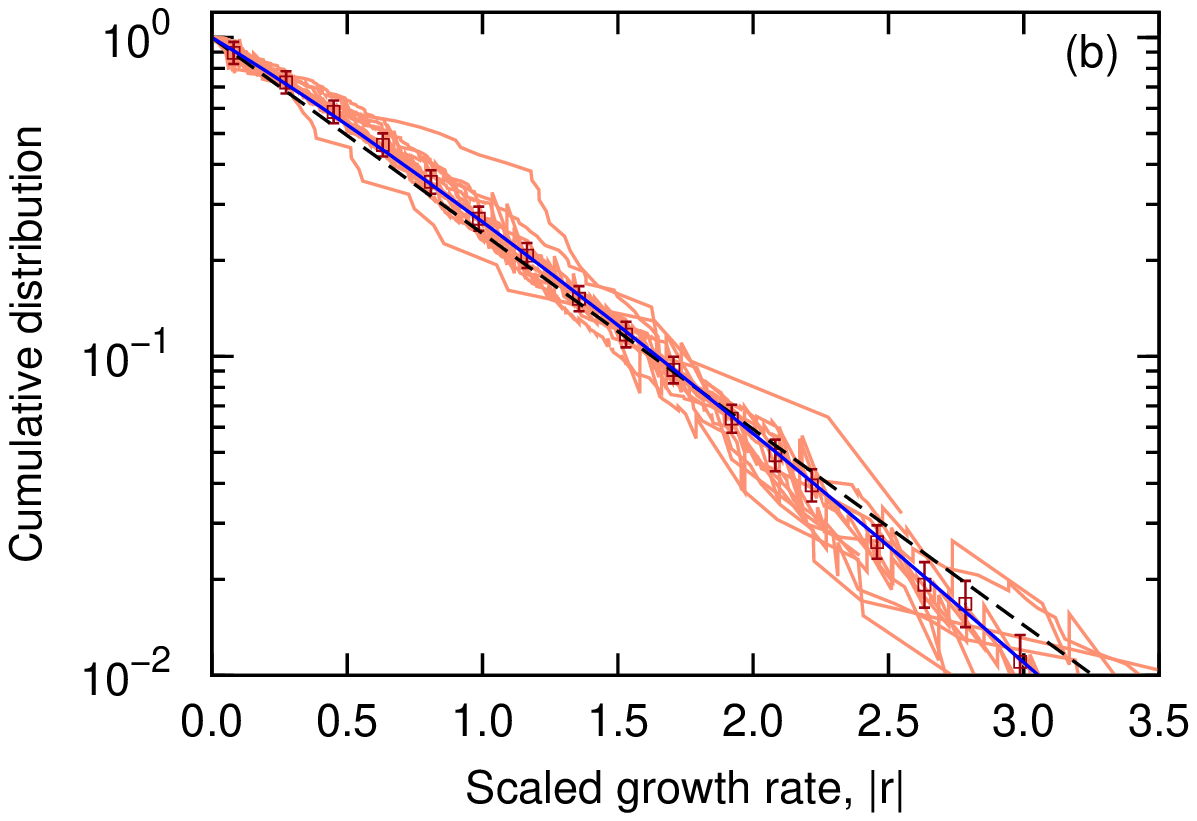}
\caption{{Normalized cumulative distribution on mono-log scale}. The {twelve continuous} red lines are the cumulative distributions of the normalized growth rates $r$, 
for each log-spaced range of initial values of $S$. The squares are the average values of the CDFs by windows in the {scaled growth rate, $r$}. The dashed lines show the Laplace distribution with unitary standard deviation and zero mean (\ref{eq:5}). The continuous lines refer to a stretched Gaussian, equation (\ref{eq:6}) with $c=1.2$. (a) Positive values of $r$ and (b) {module of the negative values of $r$}. The error bars are $95\%$ confidence intervals obtained via bootstrapping \cite{efron}.}
\label{fig:cumulative}
\end{figure}

{We divided the data in log-spaced intervals of $S$. Then,} by using the scaled variable $r$, we have verified that all curves with different ranges of $S$ collapse approximately in one single CDF (see Fig. \ref{fig:cumulative}), indicating that the distribution of growth rates exhibits the same functional form for distinct size scales. From this figure, we also note that the Laplace distribution presents small, but systematic, deviations from the empirical data.

\begin{figure}[!h]
\centering
\includegraphics[scale=0.8]{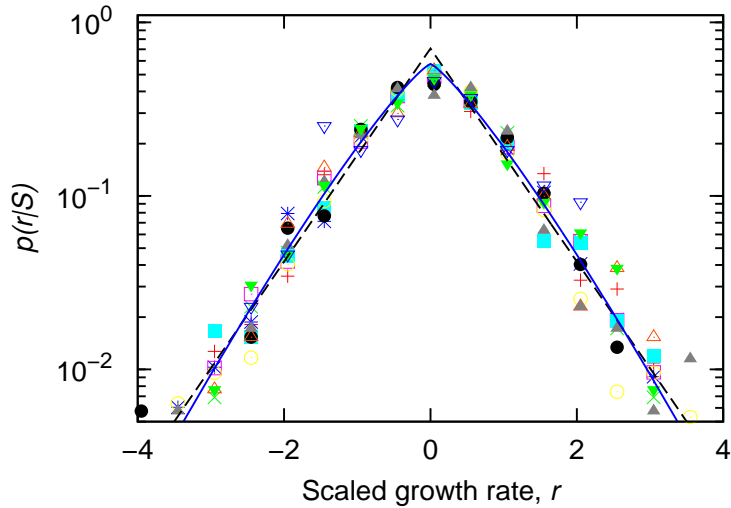}
\caption{Conditional probability density function $p(r|S)$ of the normalized annual growth rates $r$ on mono-log scale. The different symbols represent each interval of $S$ employed in the analysis of the PDF. Here, the intervals of $S$ are the same used in Fig. \ref{fig:cumulative}. The dashed black line is the Laplace distribution (\ref{eq:5}) with unitary standard deviation and zero mean. The continuous blue line is the stretched Gaussian (\ref{eq:6}) with parameter $c=1.2$ .}
\label{fig:tentshaped}
\end{figure}

In an attempt to overcome these deviations, we have fitted the data by using the one-parameter distribution

\begin{equation}
p(r|S)=\frac{c}{2}  \sqrt{\frac{\Gamma \left(3/c\right)}{\Gamma \left(1/c\right)^3}}
   \; \exp \left[-\left(\frac{\Gamma \left(3/c\right)}{\Gamma \left(1/c\right)}\right)^{c/2}
   |r|^c\right]\,.
\label{eq:6}
\end{equation}
This distribution is a ``stretched Gaussian'' ~\cite{richardson,mendes} and it can recover the standard Gaussian for $c=2$ and the Laplace distribution
for $c=1$. In our case, we have found that $c\approx1.2$ gives a more accurate description for the empirical CDFs (see Fig. \ref{fig:cumulative}).
{Note that, if we consider directly $p(r|S)$ instead of its CDF, we also obtain the tent-shaped distribution as shown in Fig. \ref{fig:tentshaped}.}

\begin{figure}[!h]
\centering
\includegraphics[scale=0.64]{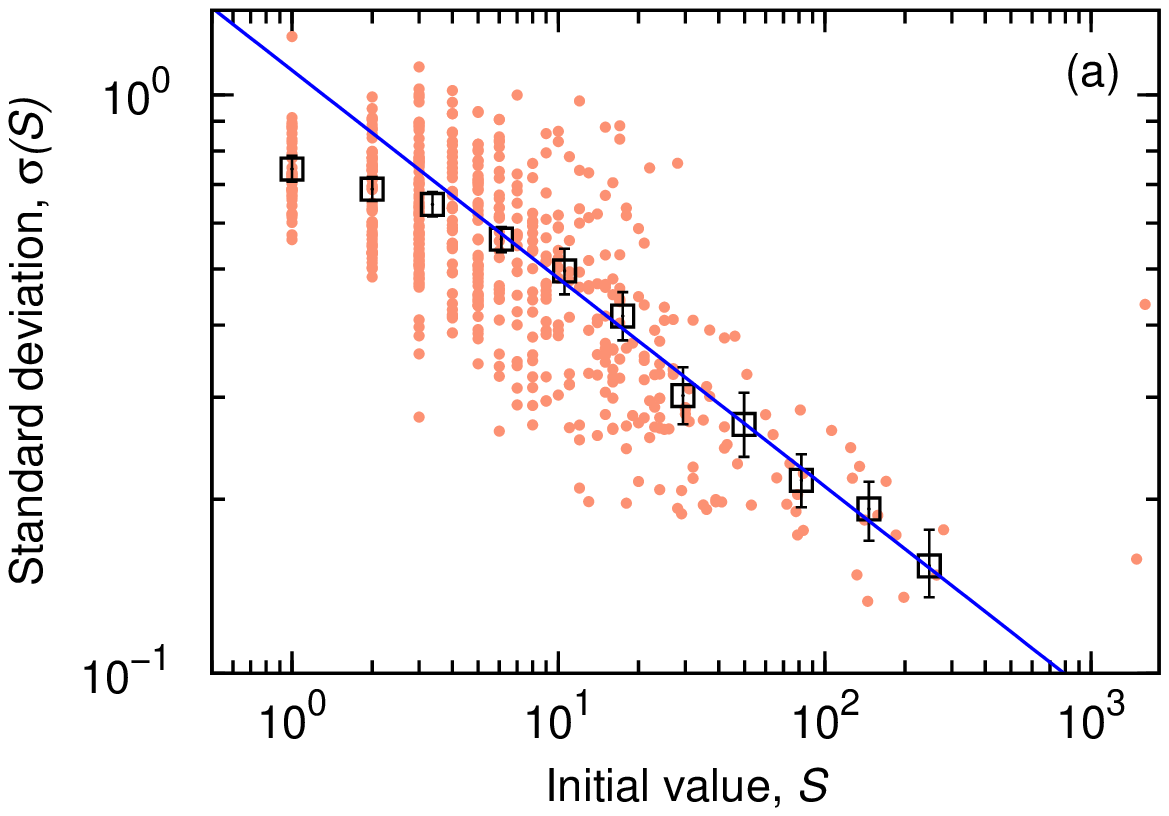}
\includegraphics[scale=0.64]{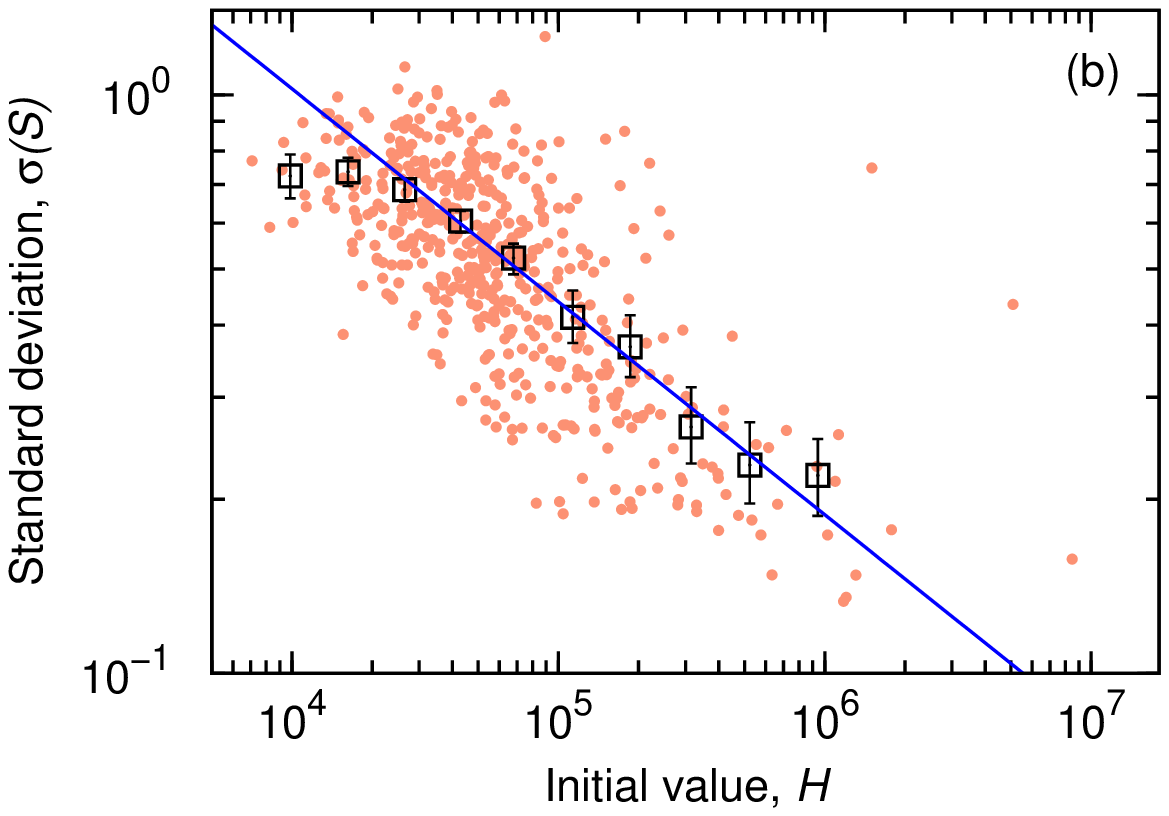}
\caption{{Standard deviation versus the city size on a log-log scale}. (a) The red points are the standard deviations of the observed annual homicide growth rates plotted against the initial value $S$ of homicide time-series for the 528 examined cities. The black squares are window averaged values of the standard deviation $\sigma$ of the growth rates $R$ as a function of the average value of $S$. (b) Here, we replace the initial value of $S$ by the population $H$ for the city. The black squares are window averaged values of the standard deviation $\sigma$ of the growth rates $R$ as a function of the average value of $H$. For both figures, the continuous lines are linear fits to the main tendencies of data on log-log scale (equation (\ref{eq:7})), yielding an $\beta_H \approx \beta_S \approx 0.36$. The error bars are $95\%$ confidence intervals obtained via bootstrapping \cite{efron}.}
\label{fig:law}
\end{figure}

Returning to Fig. \ref{fig:distribution}, we note that, for small values of $S,$ the distributions have longer tails and that the tails become shorter as $S$ increases. This result and Fig. \ref{fig:evolution} indicate that large fluctuations in the homicide growth rates are observed in small towns and that the standard deviation $\sigma$ tends to decrease with the increasing of $S$. At this point, we are considering the number of homicides as a measure of the city size since, in principle, the average number of crimes increases with the number of inhabitants {(see Fig. \ref{fig:crimepop} and \cite{lievano})}. In our analysis,  we found that the main tendency of the relationship between $\sigma$ and $S$ is well adjusted by a power-law:
\begin{equation}
\sigma  \sim  S^{-\beta_S}
\label{eq:7}
\end{equation}
with $\beta_S \approx 0.36$ for $S>8$ (see Fig. \ref{fig:law}a). However, the relation between the number of crimes and population is non-linear \cite{lievano} and not completely understood when we take all city sizes into account, { i.e., all Brazilian cities}. For this reason, we directly investigate $\sigma$ as a function of the total population $H$ of each city in 1980. As a result, we verify that $\sigma \sim  H^{-\beta_H}$ with $\beta_H \approx \beta_S$ as shown in Fig. \ref{fig:law}b. {This result is in agreement with Fig. \ref{fig:crimepop}, since when taken the 528 cities into account, the relationship between the number of homicides and the population of the cities is approximately linear, as discussed in the second paragraph of section 2.}

In addition to $R(t)$, we also investigate biannual, $R_2(t)$, and triannual, $R_3(t)$, homicide growth rates. In these last two cases, we have the same scenario of the annual growth rates. However, the numerical values of the parameters related to the annual, biannual and triannual growth rates, in general, are not equal, as shown in Table 1. Note that each $\beta_S$ value and the corresponding $\beta_H$ are very close to each other and that they decrease as the time interval increases. If we consider $\sigma$ as a function of $S(t)$ or $H(t)$ in years other than $t=1980$, $\sigma$ also presents an asymptotic power-law behavior. However, the $\beta$ exponents decrease as $t$ increases. This fact is more pronounced when we take $\sigma$ as a function of $S(t)$ than $H(t)$. In the case of $H(t)$, the exponents are almost constant. {As a last concern, we consider annual growth rates $R(t)$ of cities with at least one homicide per year in smaller periods, i.e., starting in years subsequent to 1980 and ending in 2009. Naturally, this kind of analysis contains different initial conditions. Nevertheless, the $\beta_S$ and $\beta_H$  values fluctuate around $0.33$ and $0.36$ respectively.} All these results as well as the previous ones are robust to window size change.

\begin{table}[h!]
\label{tab:summary}
\begin{center}
\caption{\bf{Summary of the results}}
\begin{tabular}{|c|c|c|c|c|}
\hline
Homicide growth rates & Distribution form & Parameter $c$ (SG) & $\beta_S$ & $\beta_H$  \\
\hline
Annual & Tent-shaped & 1.2 & 0.36 & 0.36 \\
\hline
Biannual & Tent-shaped & 1.2 & 0.29 & 0.28 \\
\hline
Triannual & Tent-shaped & 1.2 & 0.23 & 0.21 \\
\hline
\end{tabular}
\end{center}
\begin{flushleft}
A summary of our empirical results about annual, biannual and triannual logarithmic homicide growth rates for Brazilian cities is given (see equation (\ref{eq:1})) considering data from 1980 to 2009. The second and third columns inform us the shape of the collapsed distributions, where the parameter $c$ refers to the stretched Gaussian (SG) and indicates the deviation from the Laplace distribution ($c=1$) (see equation (\ref{eq:6})). The parameters $\beta_S$ and $\beta_H$ are asymptotic exponents of the main tendencies of the standard deviations $\sigma$ of the logarithmic growth rates (see equation (\ref{eq:7})).
\end{flushleft}
\end{table}

\section{Discussion}

A large part of the literature about criminality tries to relate crime rates to possible explicative variables as average income, unemployment, inequality, gender, age, education level, race and others related to delinquency, even though statistical data of crime rates cannot give much information on this \cite{gordon2}. {Certainly, cities with the same size can present different numbers of crimes (as exemplified in Fig. \ref{fig:crimepop}) usually due to socio-economic aspects. However, in our analysis we have not specified such dependences since our results arise from a more general statistical context}. This fact can be viewed as reminiscent of the concept of universality found in statistical physics, where different systems can be characterized by the same fundamental laws, independent of the microscopic details \cite{stanley}. {Systems with the same values of critical-point exponents and scaling functions are said to belong to the same universality class \cite{hstanley}. To find an exponent for the relationship between standard deviation and city size is, somehow, to indicate the class to which the system belongs. In fact, in other contexts \cite{havemann,picoli,keitt,labra,stanley,prl,fu,sergio} the analog of this exponent generally assumes different values.}

The presence of non-systematic errors on two commonly employed databases, FBI (Federal Bureau of Investigation) and UCR (Uniform Crime Report U.S.A.) ones, has been pointed out due to a lack of reports from some local agencies \cite{gordon2}. For our analysis as well as in another recent one \cite{lievano}, we choose the homicide crime to minimize this unwanted aspect of data since, in principle, any kind of demise is recorded by DATASUS \cite{datasus}.
These facts indicate that our results are robust and can be a useful guide for future works on criminality modeling.

We found that the distributions of homicide growth rates exhibit a scale-invariant structure in distinct city sizes, as shown in Fig. \ref{fig:cumulative}. Our results show that the distributions of annual, biannual and triannual homicide growth rates are close to a Laplace distribution. 
We also found that the standard deviations of growth rates have a tendency to decrease as the initial number of crimes increases (Fig. \ref{fig:evolution} and Fig. \ref{fig:distribution}). Moreover, the mean value of standard deviations are well approximated by power-laws (Fig. \ref{fig:law}). Similar results on annual growth rates and power-laws have also been reported for other complex organizations but usually with different values of $\beta$ exponents \cite{havemann,picoli,keitt,labra,stanley,prl,fu,sergio}. This parallel between the criminality with complex organizations suggest that there are  similarities in their growth dynamics.

As pointed out, there is a universality in the way the fluctuations in growth rates of homicides in cities are distributed, even though they differ in many aspects, including the details of people education, public health, cultural development and ethnicity, for example. Our results also indicate that the power-law dependence in this social system could be due to some emerging scenario as well as it occurs in the growth rate dynamics of physical and biological systems, scientific publications, economic and religion activities. Here, trying to encompass the above universal features and in a different perspective from several works that have applied a variety of techniques and models in the search for a characterization of crime behavior \cite{gordon2}, we present motivations to stochastic models proposed in the context of complex organizations in the search for some insight on the study of crime. Although modeling is not the main objective of our empirical investigations, we discuss some aspects of our results that indicate a connection between criminality and the framework of complex organizations.

We can consider that  complex organizations have an internal structure involving subunits \cite{havemann,picoli,keitt,labra,stanley,prl,fu,sergio}. Also, qualitatively consistent with criminality, for something to occur other things must take place. These facts go towards a multiplicative process for crime dynamics. Gibrat's model, initially proposed in the context of firm growth \cite{gibrat}, plays the role of a first attempt to describe the growth of complex organizations.  In our crime context, it can be written as 
$S(t+1)-S(t)= A\, \epsilon(t)\, S(t)$,
where $S(t)$ is the number of homicides in a city in the year $t$, $\epsilon(t)$ is a random number following a Gaussian distribution with zero mean and unitary standard deviation, and $A$ is a positive constant. As we can see, this model assumes that homicide growth is independent of the number of homicides and uncorrelated in time. Unfortunately, this simple approach is not enough to describe our empirical findings such as the tent-shaped distributions in the growth rates and the asymptotic power-laws in the standard deviations. However, its simplicity could be viewed as a first insight for the comprehension of mechanisms that govern the homicide growth rate. Generalizations of this model can take interaction among the subunits (e.g., non-linearity) and relevance of previous number of homicides (memory) into account. As a final remark of this brief discussion about modeling, we illustrate that, if non-linearity and memory are incorporated to Gibrat's model, a substantial improvement is obtained from it. For this illustration, we follow Ref.~\cite{picoli}, i.e., we replace $A\, \epsilon(t)\, S(t)$ by $\lambda(t)\, S^k (t)$, where $\lambda(t) = [A+B\,\lambda(t-1)]\,\epsilon(t)$ and $A$, $B$ and $k$ are positive constants ($k = 1$ and $B = 0$ recover Gibrat's model). The value of memory parameter $B$ basically dictates the degree that the probability distributions of homicide growth rates approach to tent-shapes. On the other hand, the parameter $k$ essentially fixes the $\beta_S$ value. Thus, this model is enough to reproduce two basic qualitative aspects of our annual homicide growth rate data: the tent-shaped form for the scaled distribution and the power-law behavior for the standard deviation.

\section{Conclusions}
Our study on homicide growth rates gives a new quantitative characterization of criminality. Using a framework of complex organizations, we have investigated the relation between the number of homicides and the corresponding annual, biannual and triannual logarithmic growth rates for Brazilian cities with at least one murder per year from 1980 to 2009. We have obtained two main empirical findings based on these three growth rates. One is that the standard deviations of the growth rates as a function of the number of initial murders follows asymptotic power-law decays. In addition, if the number of inhabitants is considered instead of the initial value of murders, the power-law behaviors of the standard deviations remains essentially unchanged. The other main empirical result is that the distributions of growth rates are non-Gaussian for distinct size scales but with the same shape, indicating the presence of scaling laws in the dynamics of crime growth rate. This non-Gaussian distribution is well described by the Laplace distribution, where the small corrections can be accomplished by stretched Gaussians. In a quantitative way, all these aspects are summarized in Table 1.

We report that similar empirical behaviors were also observed in a variety of other complex organizations, suggesting a universality and pervasiveness of patterns in their growth dynamics. Since this universal behavior is not dependent of details from the system, models from complex organizations are good candidates to be employed when investigating the crime dynamics. From the literature of these organizations, we consider a model that reproduces two basic qualitative aspects of our data: the non-Gaussian distributions and the power-law exponent of the standard deviations.

The quantitative investigations of the emerging behaviors go in the direction that there are some interacting structures where random multiplicative processes take place. Furthermore, since many works are still controversial due to difficulties inherent in data collection and also to technical issues in statistical treatment, our results can be a useful guide for further works on criminality modeling.

\section*{Acknowledgements}
We are grateful to CNPq and CAPES (Brazilian agencies) for financial support.

\end{document}